\documentclass[aps,prx,twocolumn,superscriptaddress,nofootinbib]{revtex4-1}
\usepackage{epsfig}
\usepackage{bm}
\usepackage{amsmath}
\usepackage{amssymb,color}
\usepackage[Symbol]{upgreek}
\usepackage{marginnote}

\begin{document}

\title{Echo-Ramsey Interferometry with Motional Quantum States}

%\author{D. Hu $^ {1}$, L. X. Niu$^ {1}$, S. J. Jin$^ {1}$, X. Z. Chen$^ {1} $ , G. J. Dong$^ {2,3}$, J. Schmiedmayer$^ {4 \ast}$, X. J. Zhou$^ {1,2 \ast\ast}$}
%\maketitle
%
%\begin{spacing}{1.5}
%
%\begin{affiliations}
%\item
%School of Electronics Engineering and Computer Science, Peking
%University, Beijing 100871, China
%$^{\ast\ast}$  E-mail: xjzhou@pku.edu.cn
%\item
%Collaborative Innovation Center of Extreme Optics, Shanxi University, Taiyuan, Shanxi 030006, China
%\item
%State Key Laboratory of Precision Spectroscopy, East China Normal
%University, Shanghai, 200062, China
%
%\item
%Vienna Center for Quantum Science and Technology, Atominstitut, TU
%Wien, Stadionallee 2, 1020 Vienna, Austria $^{\ast}$  E-mail:
%schmiedmayer@atomchip.org
%\end{affiliations}

\author{D. Hu} 
\thanks{D. Hu and L.X. Niu contributed equally to this work}
\author{L. X. Niu} 
\thanks{D. Hu and L.X. Niu contributed equally to this work}
\author{S. J. Jin} 
\author{X. Z. Chen} 
\affiliation{School of Electronics Engineering and Computer Science, Peking University, Beijing 100871, China}
\author{G. J. Dong} 
\affiliation{Collaborative Innovation Center of Extreme Optics, Shanxi University, Taiyuan, Shanxi 030006, China}
\affiliation{State Key Laboratory of Precision Spectroscopy, East China Normal University, Shanghai, 200062, China}
\author{J. Schmiedmayer} 
\email{schmiedmayer@atomchip.org}
\affiliation{Vienna Center for Quantum Science and Technology, Atominstitut, TU Wien, Stadionallee 2, 1020 Vienna, Austria}
\author{X. J. Zhou}
\email{xjzhou@pku.edu.cn}
\affiliation{School of Electronics Engineering and Computer Science, Peking University, Beijing 100871, China}
\affiliation{Collaborative Innovation Center of Extreme Optics, Shanxi University, Taiyuan, Shanxi 030006, China}

\date\today

\pacs{pacs}

\begin{abstract} 
Ramsey interferometers (RIs) using internal electronic or nuclear states find wide applications in science and engineering. 
We develop a matter wave Ramsey interferometer for \emph{motional} quantum states exploiting the S- and D-bands of an optical lattice and identify the different de-phasing and de-coherence mechanisms. We implement a band echo technique, employing repeated $\pi$-pulses.  This suppresses the de-phasing evolution and significantly increase the coherence time of the motional state interferometer by one order of magnitude. We identify thermal fluctuations as the main mechanism for the remaining decay contrast. 
Our demonstration of an echo-Ramsey interferometer with motional quantum states in an optical lattice has potential application in the study of quantum many body lattice dynamics, and motional qubits manipulation.
\end{abstract}

%One of its central requirements for further development is to greatly extend the coherence time of the motional quantum state. 
%Recently, motional RIs exploiting the interference of quantum states are emerging and have the potential to be applied in the study of many-body physics and quantum information with motional states.

\maketitle

Ramsey interferometers (RIs) using internal electronic or nuclear states~\cite{Ramsey} already have played an important role in accurate quantum state engineering and quantum metrology, such as in nuclear magnetic resonance~\cite{NMR}, atomic clocks~\cite{Aclock}, quantum information~\cite{Qinf} and quantum simulation~\cite{Qsimu}. In general, echo techniques are used in RIs to suppress dephasing for significantly increasing the coherent time~\cite{Schmiedmayer1,nmr2,Bo}. Even though motional states of particles are on the same footing as internal states in current quantum technologies, i.e., motional qubits~\cite{Neg}, motional logic gates~\cite{Lee,Wineland,Briegel}, motional quantum error correction~\cite{Steinbach}, quantum entanglement and coherent control~\cite{Sangouard, Zhuang}, quantum metrology with nonclassical motional states~\cite{Dalvit}, and communication multiplexing~\cite{Muga}, conventional echo-RIs rarely exploit the quantum interference of external center-of-mass motional states. 

Recently, a RI with motional states of a Bose-Einstein condensate (BEC) trapped in an anharmonic potential has been demonstrated with $92\%$ contrast for several cycles~\cite{Schmiedmayer2}. This proof-of-principle experiment holds great promises for studying quantum many-body physics out of equilibrium, quantum metrology with non-classical motional states and quantum information processing with motional qubits. Recently this new technique has been used to investigate decoherence and relaxation dynamics~\cite{cohe}, or to measure the phononic Lamb shift~\cite{Lamb}.

Ultracold atoms trapped in an optical lattice (OL) is an ideal test platform for studying quantum many-body dynamics \cite{Bloch}, and have also been widely used as a high-precision metrology tool~\cite{Derevianko}. Conventionally, these atoms are prepared in the lowest band of the OL. Over last few years, there is an increasing experimental and theoretical interest to prepare atoms in higher bands~\cite{Bloch2,Schori}, opening a new way to simulate exotic orbital physics in strongly correlated matter with rich degrees of freedom, i.e., the formation of multi-flavor systems~\cite{flav}, supersolid quantum phases in cubic lattices~\cite{Larson}, quantum stripe ordering in triangular lattices~\cite{stripe} and Wigner crystallization in honeycomb lattices~\cite{Wig}.

In this paper, we demonstrate an echo-RI with  motional quantum states (MQS) of atoms employing the S- and D-bands in an OL. Due to the lack of selection rules for lattice band transition, a key challenge for constructing this RI is to realize $\pi$- and $\pi/2$-pulses analogous to those in conventional RIs. Using a shortcut loading method~\cite{FF, Chen}, we have designed sequences of optical pulses~\cite{Liu,Zhai}, analogous to a $\pi$- or $\pi/2$-pulse, to efficiently prepare a superposition of atoms in S- and D-band states in the OL at zero quasi-momentum with high fidelity within tens of microseconds, which is much shorter than the characteristic time scales of
the decay process. Keeping the lattice on, we observe state interference and measure the decay of the coherent oscillations. 

We have identified the mechanisms leading to the RI contrast reduction as following: the nonuniform optical lattice depth, interaction-induced transverse expansion after loading the atoms from the harmonic trap into the optical lattice, laser intensity fluctuation and thermal fluctuations at finite temperature. We then implement a matter-wave band echo technique to significantly suppress all the contrast decay effects except for quantum and thermal fluctuations, increasing the coherence time to 14.5 ms compared to 1.3 ms without echo at condensate temperature of 50 nK and $10E_\mathrm{r}$ lattice depth. 

\section{A Ramsey interferometer within an optical lattice}

\subsection{Experimental implementation} 
\begin{figure}
\begin{center}
\includegraphics[width=1.0\linewidth]{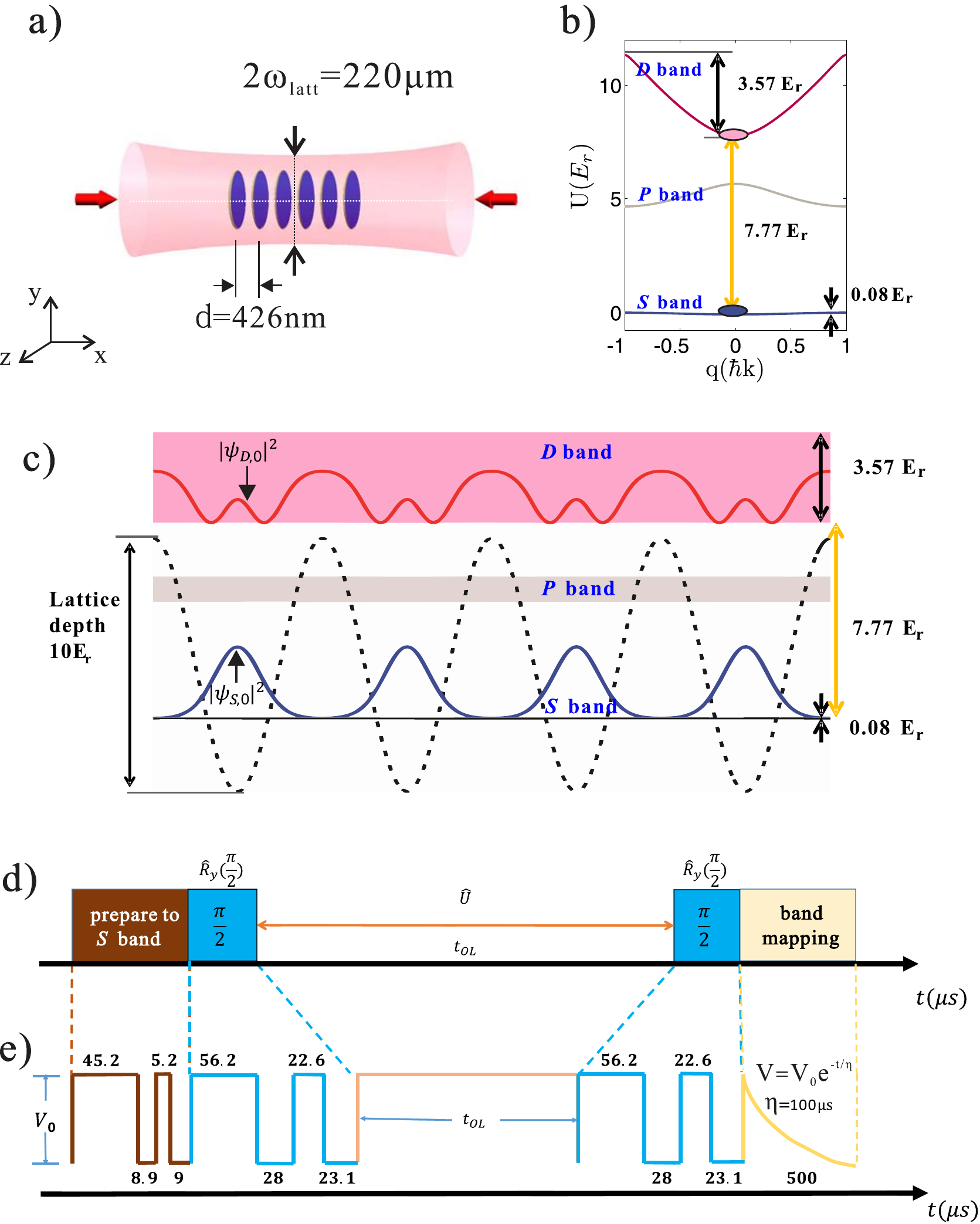}
\end{center}
\caption{\textbf{Experimental configuration} for a Ramsey Interferometer in a $V_0=10E\mathrm{r}$ lattice:  
\textbf{a)} The BEC is divided into discrete pancakes in $yz$ plane by an 1-dimensional optical lattice along $x$ axis with a lattice constant $d=426$ nm. 
\textbf{ b)} Band energies for the S-band and the D-band. 
\textbf{c)} Wave-functions for the S-band and the D-band. At zero quasi-momentum, the widths of the bands are $0.08E_\mathrm{r}$ and $3.57 E_\mathrm{r}$, respectively. The band gap between S and D is $7.77E_\mathrm{r}$. 
\textbf{d)} Time sequences for the Ramsey interferometry. The atoms are first loaded into the S band of OL, followed by the RI sequence:  $\pi/2$ pulse,  holding time $t_{OL}$, and the second $\pi/2$ pulse. Finally band mapping is used to detect the atom number in the different bands.  
\textbf{e)} The used pulse sequences designed by an optimised shortcut method.}
\label{f1}
\end{figure}

Our experiment start with a BEC of $^{87}Rb$ prepared in a hybrid trap formed by a single-beam optical dipole trap with wavelength 1064 nm and a
quadruple magnetic trap (Details are presented in our former works~\cite{Wang, Hu}). A nearly pure condensate of about $1.0\times 10^ {5}$ atoms at the temperature 50 nK is achieved with the harmonic trapping frequencies $(\omega _{x},\omega _{y},\omega _{z})=2\pi \times $(24, 48, 58) Hz, respectively. The system's temperature can be controlled by the evaporative cooling process (see appendix). After preparation of the condensate, a one dimensional optical lattice is formed by two counter-propagating laser beams with wavelength 852 nm resulting in a  lattice constant $d=426$ nm along $x$ axis, as shown in Fig.~\ref{f1}(a). Using a shortcut control method~\cite{Liu}, the BEC is loaded into the lowest lattice band with the quasi-momentum $q=0$ within a few microseconds with nearly $100\%$ fidelity. The interaction energy of the condensate in the ground state of OL with a lattice depth of $10E_\mathrm{r}$ is 1.2 kHz ($E_\mathrm{r}=(\hbar k)^ {2}/2m$ is the recoil energy, $k=2 \pi/426$ nm$^{-1}$ is the wave vector associated with the lattice and $m$ is the atomic mass).

In the experiment, we construct the Ramsey interferometer with Bloch states $\phi_{i,q}$ with energy $\epsilon_{i,q}$. We use the lowest band $\phi_{S,0}$ and the second excited band $\phi_{D,0}$ at quasi-momentum $q=0$, denoted as $|S\rangle,|D\rangle$ respectively in the following, to form a superposition state $\psi=a_S|S\rangle+a_D|D\rangle$. The two bands of S and D are considered a \emph{two-state} system (spin-1/2 system), i.e., the two states can be expressed as $\binom{a_S}{a_D}$. The scheme for the band energy and the superposition states are shown in Fig.~\ref{f1}(b) and Fig.~\ref{f1}(c), respectively, with the lattice depth $V_0=10.0E_\mathrm{r}$. During the lattice pulse sequence, a accusto-optic modulator controlled by an RF switch is used to switch on and off the lattice potential (light) quickly.

Manipulating the pseudo-spin system has its own challenges: unlike conventional RI where selection rules can be used to prepare population in two sates, the lattice band transition, similar to transition for vibration states in molecules~\cite{mole}, has no selection rules. Thus a $\pi/2$-pulse analogous to those in conventional RIs had to be numerically designed, so that the atoms in S-band and D-band are to be transferred into the target states $|\psi_{1}\rangle$=$(|S\rangle+|D\rangle)/\sqrt{2}$ and $|\psi_{2}\rangle$=$(-|S\rangle+|D\rangle)/\sqrt{2}$, respectively.  Extending our optimal control method~\cite{Liu, Zhai}, we construct a $\pi/2$ pulse from a sequence of lattice pulses with special selected timing and duration as shown in Fig.~\ref{f1}(e).  We achieve a fidelity of $98.5\%$ and $98.0\%$, for atoms initially on the S- and D-band respectively. (See appendix).

The full time sequence for RI is shown in Fig.~\ref{f1}(d). First the atoms in the harmonic trap are transferred into the S band by a fast loading process, then the first pulse $\hat R(\pi/ 2)$ is applied to prepare an initial superposition state $\hat R(\pi/2)\binom{1}{0}=\frac{1}{\sqrt{2}}\binom{1}{1}$. The final state, after evolution in the OL for time $t_{OL}$ and a second $\pi/2$ pulse can be expressed as:
\begin{equation}\label{e1}
    \psi_f=\hat R(\pi/2) \hat U(t_{OL}) \hat R(\pi/2)\psi_i,
\end{equation}
with $\hat R(\alpha)=(\cos\frac{\alpha}{2}-i\sin\frac{\alpha}{2}) \hat \sigma_y$, and the evolution operator $\hat U(t)=(\cos\omega t+i\sin\omega t)
\hat \sigma_z$, here $\omega=(\epsilon_{D,0}-\epsilon_{S,0})/\hbar$ is the frequency difference between two states, and $\hat\sigma$ the Pauli matrix, $\alpha$ denotes rotation angle around an axis in the Bloch sphere.

We then apply band mapping~\cite{Bloch2} to read out the state of the RI.  For band mapping the lattice depth is exponentially-ramped down in the form $e^{-t/\eta}$ with a  characteristic decay time $\eta=100\mu s$ for a total length of 500 $\mu$s. Then absorption imaging is used to measure the population in the different bands after 31 ms time of flight (TOF) (see appendix). The atoms originally occupying $|S\rangle$ state populate a narrow Gaussian distribution around $0\hbar k$.  The atoms originally occupying the  $|D\rangle$ are detected at $\pm 2\hbar k$ in the side zone. The total atom number we detected consists of condensed atoms and  thermal atoms. From a bimodal fitting to each peak~\cite{Schori}, we can quantitatively determine the population of condensate, and we note the condensed atom number as $N_\mathrm{S}$ ($N_\mathrm{D}$) for S-band (D-band) (see appendix).

\subsection{Ramsey Interferometer in an optical lattice}

\begin{figure}
\begin{center}
\includegraphics[width=1.0\linewidth]{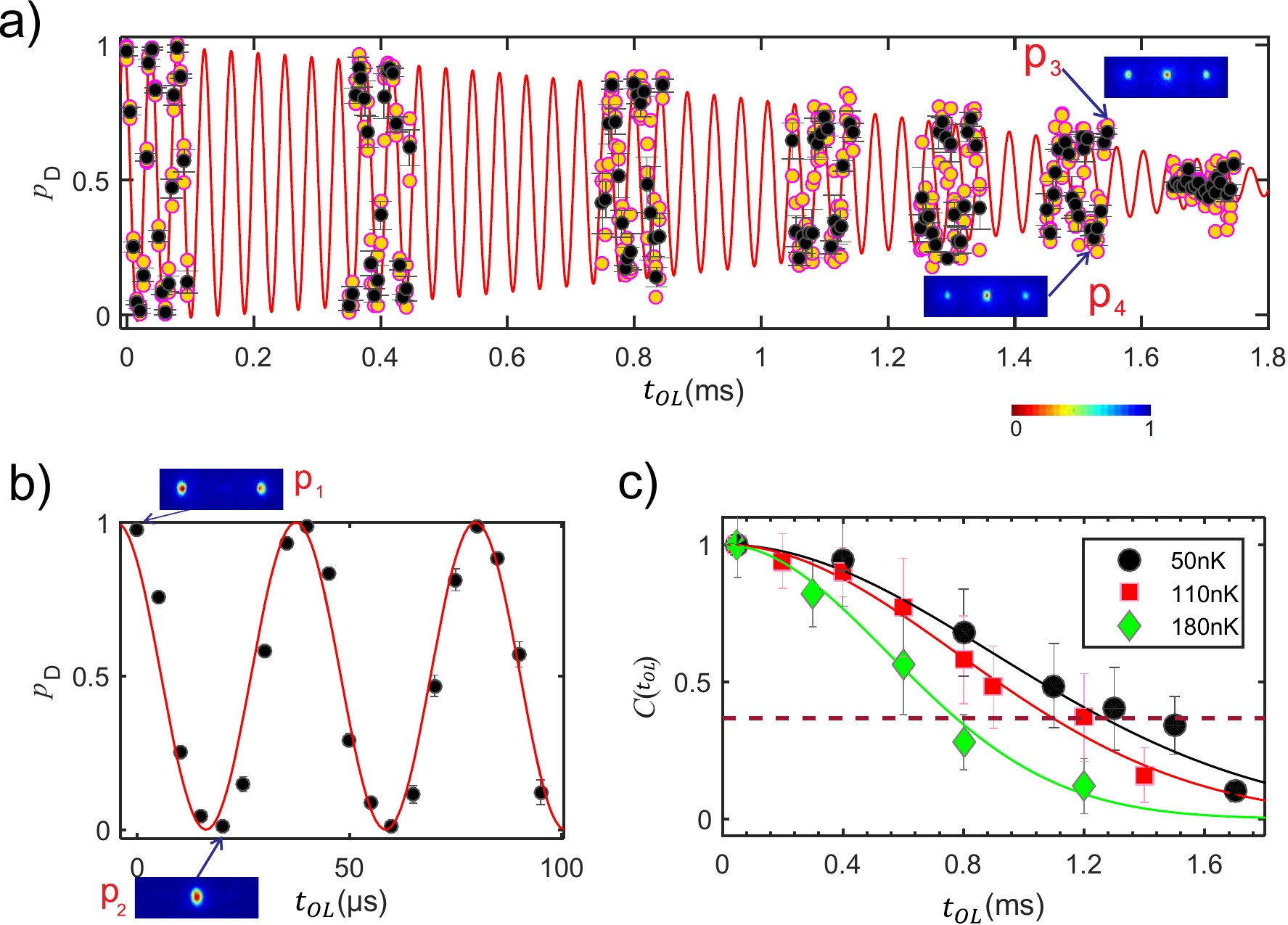}
\end{center}
\caption{\textbf{Signal for a Ramsey-interferometer} between the S- and D-bands for $V_{0}=10E_\mathrm{r}$. 
\textbf{a)} The oscillation of the population of atoms in the D-band $p_{\mathrm{D}}$ (initial temperature $T=50$ nK). The dots with errorbars are the mean of three measurements, while single measurement is shown as red circle, and the red solid line is the corresponding fitting curve to the mean value.  The inserts show typical time of flight pictures after band mapping.
\textbf{b) }Detail of the oscillation for $t_{OL}<100$ $\mu$s (shown as black dots).
Insets: The TOF images $P_{1}$ at 0 $\mu$s and $P_{2}$ at 20 $\mu$s, also the images $P_{3}$ and $P_{4}$ around $1.5 ms$. 
\textbf{c)} The decay of the interferometer contrast $C(t_{OL})$ for different temperatures.  The black dots, red squares and green diamonds corresponding to 50 nK, 110 nK and 180 nK, respectively, with the corresponding solid lines a guide to eye. The horizontal dashed line indicates the contrast drops to the value of $1/e$ that gives the coherence time $\tau$.}
\label{f2}
\end{figure}

Experimental results on the time evolution of the atom population in the D-band $p_{D}(t_{OL})=N_{D}/(N_{S}+N_{D})$  with evolution time in the lattice $t_{OL}$ are shown in Fig.~\ref{f2}(a) for the time sequence $\hat R(\pi/2)-\hat U(t_{OL})-\hat R(\pi/2)$, $V_{0}=10E_\mathrm{r}$ and $T=50$ nK. Each solid point with error bar is the mean of three measurements, shown as red circles. The red solid line is a fit of an damped oscillation with a period of $41.1 \pm 1.0 \mu s$  which is consistent to the reciprocal of the band gap energy of about $40.8 \mu s $.  The details of the early RI oscillation for $t_{OL}<100\mu s$ is shown in Fig.~\ref{f2}(b), displaying a nearly perfect oscillation between the S-band and the D-band with amplitude close to 1. At $P_{1}$ nearly all the atoms are transmitted from the S-band to the D-band by two $\pi/2$ pulses which can be seen as a $\pi$ pulse. At $P_{2}$, the two $\pi/2$ pulses offset each other due to phase evolution of two states in the lattice, and the atoms are transferred to the S band. However, when $t_{OL}$ gets longer, the oscillation amplitude decays, as shown in Fig.~\ref{f2}(a). The contrast $C(t_{OL})$ at $t_{OL}$ can be obtained by fitting the amplitude of oscillation $p_D(t_{OL})$ in Fig.~\ref{f2}(a) with
\begin{equation} \label{e2}
p_D(t_{OL})=[1+C(t_{OL})\cos (\omega t_{OL}+\phi)]/2,
\end{equation}

Fig.~\ref{f2}(c) shows the measured contrast decay versus time $t_{OL}$ for the different initial temperatures of condensates, where each contrast value is fitted from about 20 experimental points with a time step 5$\mu$s and each point is the mean value of three measurements. The error-bars are given by $95\%$ confidence bounds. The horizontal dashed line in Fig.~\ref{f2}(c) indicates the contrast drops to the value of $1/e$, which is used to define a coherence time $\tau$. This time decreases from 1.3 ms to 0.8 ms when the temperature increases from 50 nK to 180 nK.

\subsection{Contrast decay mechanisms }

\begin{figure}
\begin{center}
\includegraphics[width=1.0\linewidth]{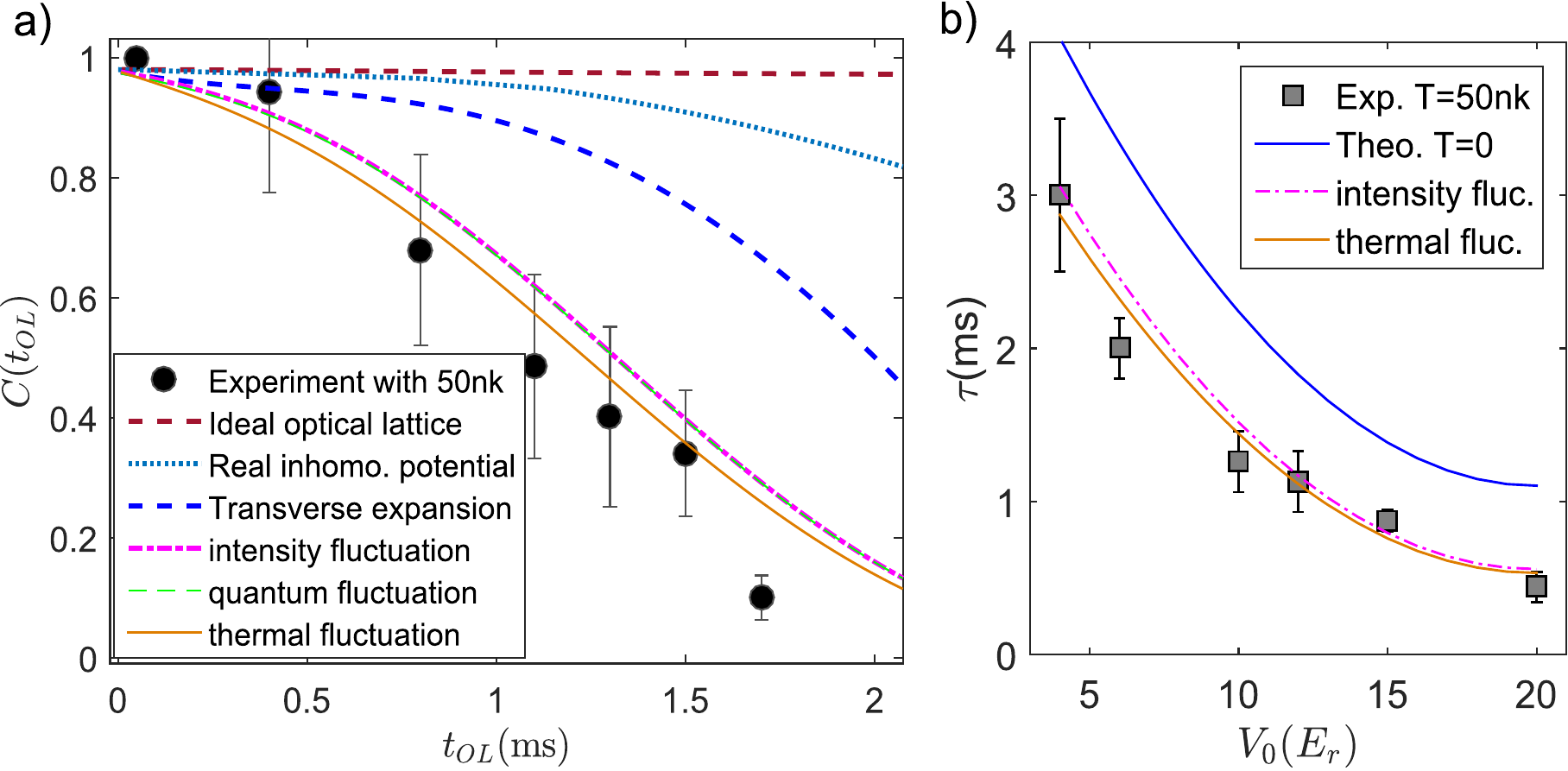}
\end{center}
\caption{\textbf{Contribution to contrast decay} for $\pi/2$-$\pi/2$ RI. \textbf{a)} The experimental contrast and the theoretical calculation for $V_0=10E_\mathrm{r}$.  \textbf{b)} The coherence time versus the different depths $V_0$. Components as described in the text.}
\label{f3}
\end{figure}

In order to improve the performance of the RI, we now investigate the mechanisms that lead to RI signal attenuation. We first study the effect of imperfect design of the $\pi/2$ pulse on the fidelity by solving Schr{\"o}dinger equation with an uniform lattice potential, and the unbalanced population between the S- and D-bands (see appendix). The numerical result (brown dashed line in Fig.~\ref{f3}(a)) shows that the imperfectly designing of $\pi/2$ pulses and the unbalanced population have negligible effects on the contrast decay during the evolution in lattice potential.

In our further theoretical analysis, we replace the ideal lattice potential by a non-uniform potential distribution to account for the Gaussian beam in the radial direction, as shown in Fig.~\ref{f1}(a), and include the quasi-momentum distribution for the condensate distributed in the harmonic trap~\cite{dep}. These two effects result in inhomogeneous broadening of the transition frequency $\omega$ between the S- and D-bands, and lead to to de-phasing and contrast decay. By solving the zero-temperature Gross-Pitaevskii equation(GPE) with the real inhomogeneous potential~\cite{Tdamping} (see appendix), we obtained the contrast as shown by the blue dotted line in Fig.~\ref{f3}(a).

The atom-atom interaction leads to transverse expansion after the fast (non-adiabatic) loading of the atoms from the harmonic trap into the 1-d optical lattice. This expansion leads to a significant reduction on contrast (blue dashed line in Fig. 3(a)).

Moreover, adding a $0.1\%$ laser intensity fluctuation further reduces the contrast as shown in the purple dash-dotted line in Fig.~\ref{f3}(a). Quantum fluctuations at zero temperature is further added using a truncated Wigner method (see appendix) and the related results shown in green dashed line nearly overlap with the dash-dotted line. In our experiment, the atom number is high enough such that the influence by the quantum fluctuation is not significant.

Finally, we take into account the thermal fluctuations using a finite temperature truncated Wigner calcualtion (see appendix). The result is shown in the orange solid line in Fig.~\ref{f3}(a), agreeing well with the experimental results shown in black dots.

The relation of the coherence time to lattice depth is presented in Fig.~\ref{f3}(b). When the lattice depth increases, the confinement in x axis gets tighter, accordingly, the effects of the expansion in radial direction will increase due to the increased interaction energy. In addition, the non-uniformity of the lattice potential (the difference of the lattice potential from center to edge) also increases with the increasing of lattice depth. Due to these two effects, the coherence time $\tau$ decreases with increasing lattice depth. The theoretical curve in Fig.~\ref{f3}(b) including all the decay mechanisms (orange solid line) fits well with an experimental measurement.

\section{An Echo-Ramsey Interferometer with  motional quantum states of atoms}

\begin{figure}[t]
\begin{center}
\includegraphics[width=1.0\linewidth]{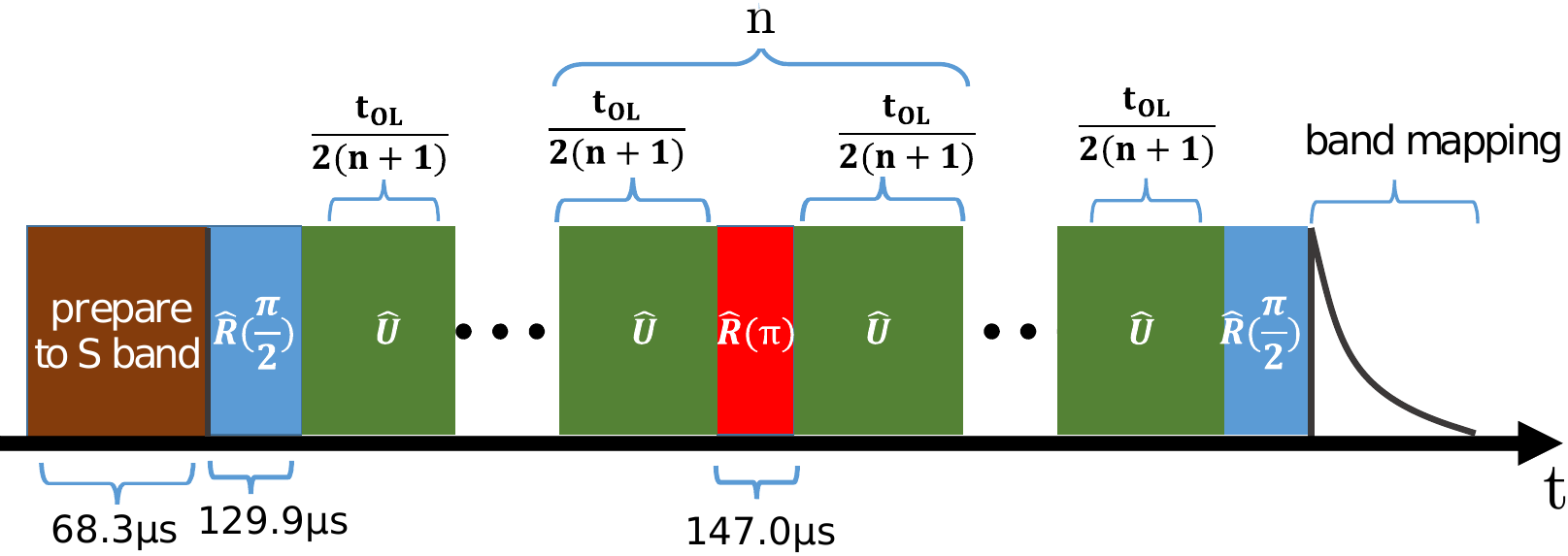}
\end{center}
\caption{\textbf{ Scheme for an echo-Ramsey interferometer}: Between two $\pi/2$ pulses, $n\times\pi$ pulse are inserted. Band mapping is implemented after the second $\pi/2$ pulse to detect the population in the different bands.}
\label{f4}
\end{figure}

\begin{figure}[bt]
\begin{center}
\includegraphics[width=1.0\linewidth]{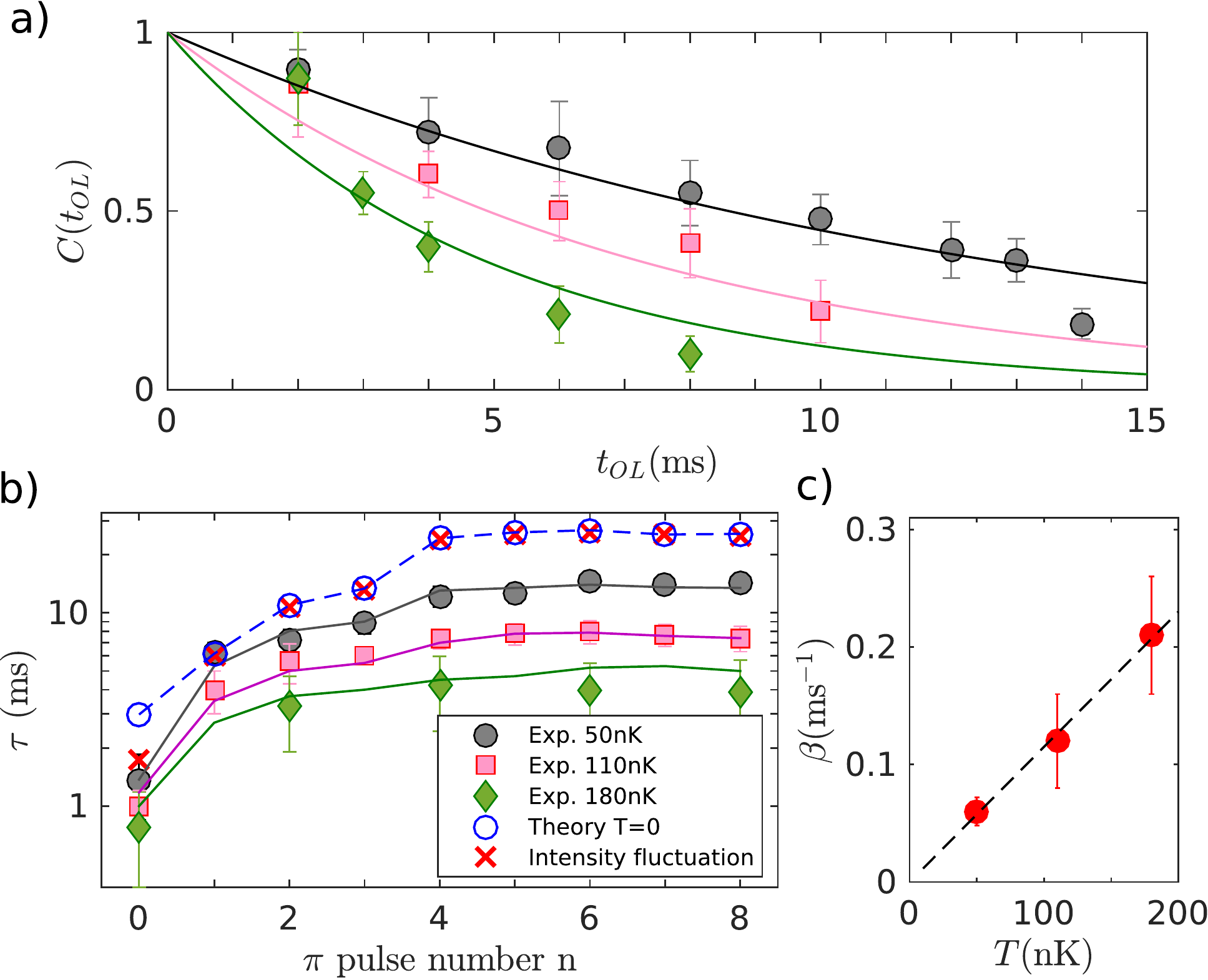}
\end{center}
\caption{\textbf{Echo-Ramsey Interferometer}:
\textbf{a)} Contrast $C$ vs  holding time $t_{OL}$ in the OL after using four $\pi$ pulses. Solid curves are a fit with an exponential function to extract the coherence time $\tau$.
\textbf{b)} Coherence time $\tau$ vs the number of applied $\pi$ pulse $n$. The circle and cross are theoretical results at zero temperature with or without intensity fluctuations respectively. Experimental results and fitting curves with parameter $\beta$ for the different temperatures are also shown. 
\textbf{c) }Experimental (red dots) decay rate $\beta$ induced by thermal fluctuations $vs$ temperature. The error-bars give a 95$\%$ confidence interval. The dashed line illustrates that the decay rate $\beta$ is proportion to the temperature. }
\label{f5}
\end{figure}

To further improve the contrast of the RI with MQS, we develop a matter-wave band echo technique. A $\pi$ pulse is designed which swaps the atom population in  $|S\rangle$ and $|D\rangle$ band. The pulse schema of our echo-RI scheme is shown in Fig.~\ref{f4}. In between the initial and final $\pi/2$ pulses we insert $n$ identical $\pi$ pulses. A single $\pi$ pulse is realised with the operation $\hat U(t_{OL}/2n) \hat R(\pi) \hat U(t_{OL}/2n)$. The full Echo-Ramsey pulse sequence is then: $\hat R(\pi/2)[\hat U(t_{OL}/2n) \hat R(\pi)\hat U(t_{OL}/2n)]^{n} \hat R(\pi/2)$. The contrast decay $C(t_{OL})$ of a $4\times \pi$ pulse echo-RI versus the holding time at various temperatures between 50 nK  and 180 nK is shown in Fig.~\ref{f5}(a).   The solid lines are fits with an exponential function which allow to extract the coherence time $\tau$.

The relationship between the coherence time $\tau$ and the number of $\pi $ pulses $n$ is shown in Fig.~\ref{f5}(b) for three different temperatures ($T$=50, 110 and 180 nK). The zero-temperature theoretical results of the coherence time can again be obtained by solving the GPE including the non-uniform lattice potential and the radial expansion, as shown by the circles in Fig.~\ref{f5}(b).
When further taking into account of the laser intensity fluctuation, the calculated coherence time is shown with the cross points.
After implementing one $\pi$ pulse, the two theoretical points merger together, which shows that the intensity fluctuation can be well suppressed with one $\pi$ pulse;
however, in order to eliminate the effects of non-uniform lattice potential and transverse expansion, more echo pulses are needed; after applying $n=6$ echo pulses,
the theoretical curves get nearly flat, showing that the two effects are well suppressed. There remains a significant discrepancy between the experimental and the zero temperature theoretical calcualtions, which we attribute to thermal fluctuations.  Lowering the condensate temperature, the thermal fluctuations get weaker, and the coherence time can be greatly increased with multiple echo-pulses. The measured coherence time $\tau$ for the RI at $T=50$ nK is 14.5 ms, which is one order of magnitude longer than the case without echo-pulses.

We will now look at the remaining fluctuations that cannot be suppressed by the echo technique:  We fit the experimental contrast with $C_{0}(t_{OL})\exp[-\beta t_{OL}]$, where $C_{0}$  is the calculated contrast for zero temperature and the decay rate $\beta$ characterises the additional decay. For different temperatures, the fitted values of $\beta$ are shown in Fig.~\ref{f5}(c) with red dots, showing a linear dependence on the temperature. This is a strong indication that the remaining decay of the interference contrast of the echo-Ramsey interferometer is caused by thermal fluctuations. To confirm this we conducted finite temperature truncated Winger calcualtions~\cite{Temperature,Temperature2} (see appendix)

\begin{figure}
\begin{center}
\includegraphics[width=1.0\linewidth]{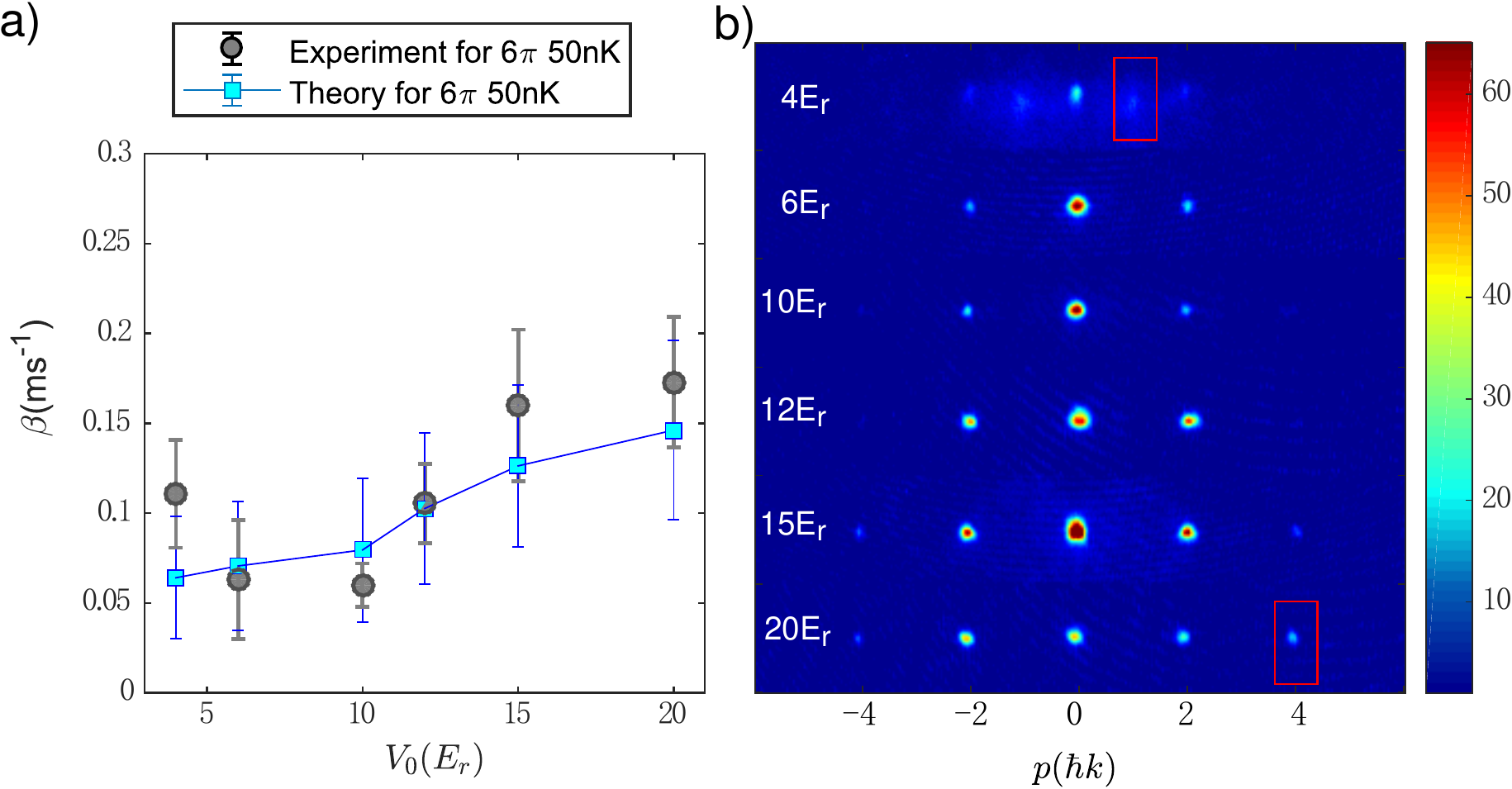}
\end{center}
\caption{\textbf{a)} The contrast decay rate from thermal effect $\beta$ versus the different lattice depth $V_0$ with six $\pi$ pulses and $T$ = 50 nK with the experimental results (solid points) and theoretical calculation at zero temperature (square points). \textbf{b)} Typical TOF images for different depths and the red square shows atom on other momentum which causes an increasing of decay rate in experiment.}
\label{f6}
\end{figure}

Finally we look at the performance of the echo-RI for different lattice depth. As shown in Fig.~\ref{f6}(a) the contrast decay rate strongly depends on the lattice depth for a condensate at 50 nK and using a sequence with six $\pi$ pulses.  The decay rate is small at medium optical lattice depth, and it becomes bigger for the shallow ($<$5$E_\mathrm{r}$) and deep lattices ($>$10$E_\mathrm{r}$). 
For the case of $V_0=4E_\mathrm{r}$, the separation between D and P bands is small and it is hard to form a close cycle between S- and D-band without any transition to P-band in practice.
We should note that for the two pulse RI, transition to other bands are not so evident due to the short lattice time. When the echo technique is applied, the coherence time is much increased, and transition to other bands induced by atom-atom interaction become important.  This can be seen in Fig.~\ref{f6}(b) where the momentum distribution for 4$E_\mathrm{r}$ has components (marked with a red square) different from $\pm 2\hbar k$.

When the lattice potential gets deeper, the band separation increases, and the coherence time increases accordingly. However, when the lattice depth is more than 10$E_\mathrm{r}$, the coherence time decreases quickly. In fact the probability for atoms to be excited to higher momentum states by the pulse is proportional to the lattice  depth. For a very deep lattice the atoms can easily obtain higher momenta \cite{YYZ}.  In Fig.~\ref{f6}(b) the momentum distributions clearly show the components at $ 4\hbar k$ for 20 $E_\mathrm{r}$. Since the Rabi transition is designed for the transition between the S- and D-band at zero quasi-momentum, the higher states will lead to a reduced pulse fidelity and a faster decay of the contrast.

%The dynamic behaviour is controlled by the lattice depth, thus, the lattice depth also has strong influence on the performance of the echo-RI,
%The experimental results are shown with the solid points, and the theoretical calculations taking the finite temperature effects into account are shown as the square points.
%In this situation, more momentum states are included, thus the high quality self-imaging (the atoms jumps back to zero momentum state)~\cite{mole2} cannot be obtained and the pulse fidelity will drop.

\section{Summary and Discussion}

In summary, we have demonstrated a Ramsey interferometer for atoms within an OL.  We further employ a matter wave band echo
technique to significantly enhance the coherence time by one order of magnitude. We have identified the mechanisms leading to the contrast
decay for the RI signal. The contrast decay is closely related to nonuniform optical lattice, laser intensity fluctuation and interaction-induced transverse expansion, and finite temperature dynamics of a BEC. All except for the effects of finite temperature can be suppressed by a matter-wave band echo sequence. Thus the damping from thermal fluctuations is well uncovered in this way.

So far, the $\pi$ pulses and $\pi/2$ pulses for echo-RI using MQS are designed based on zero temperature single atom dynamics. In furure developments,  it would be interesting to unveil quantum many body dynamics in OL, like loop structures and swallowtails~\cite{nband1,nband2,nband3}.  These deliberate quantum control technologies could be applied in quantum information and precise measurements based on MQS of atoms.

\vspace{5mm}
\section*{Acknowledgement:} 
We thank Z. K. Chen for the calculation of pulse time sequences and discussion. Thank P. Zhang, Q. Zhou, and B. Wu for helpful discussion. 
This work is supported by the National Key Research and Development Program of China (Grant No. 2016YFA0301501), and the NSFC (Grants No.11334001, No.61475007 and No. 61727819). 
G. J. Dong acknowledges the support by the National Science Foundation of China (Grants No. 11574085 and No. 91536218), and 111 Project ( B12024).  
JS acknowledges support by the European Research Council, ERC-AdG \emph{QuantumRelax}. 

%\vspace{2mm}
%\noindent Correspondence should be addressed to: \\
%Xiaoji Zhou (xjzhou@pku.edu.cn) and \\
%J. Schmiedmayer (schmiedmayer@atomchip.org).
%
%\vspace{2mm}
%\noindent D. Hu and L.X. Niu contributed equally to this work.

%\newpage

\appendix

%\subsection{ Experimental setup and detail of band mapping}
%The condensates at different temperatures have been prepared in a hybrid trap formed by an optical dipole trap and a magnetic trap by controlling evaporative cooling processing. A nearly pure condensate of $1.0\times 10^5$ atoms at 50 $nK$ is achieved with trapping frequency $2\pi\times$(24,48,58)Hz.
% A condensate of $1.1\times 10^5$ atoms at 110 nK is obtained with the trapping
%frequencies of $2\pi\times$(28,55,65)Hz with a condensate
%fraction  $50\%$. A condensate of $1.2\times 10^5$ atoms at 180 nK has been prepared with trapping frequencies of $2\pi\times$(28,80,90)Hz with $22\%$ condensate population. In the experiment, we apply a band mapping technique to detect population in different bands~\cite{Bloch2}, where we use a program-controlled attenuator (ZX73-2500-S+) between the RF source and RF switch to generate an exponential decay of the lattice depth, i.e, $e^{-t/\eta}$ with $\eta=100\mu s$.

\section{Design of Pi/2- and Pi-pulses}

In a 1D optical lattice along x-direction, the eigen-states of the atom, i.e. the Bloch states, can be expressed as the superposition of momentum states 
$$|i,q\rangle =\sum_{l=-\infty }^{+\infty }c_{i,l}(q)|2l\hbar k+q\rangle ,$$ 
where $|2l\hbar k+q\rangle$ is the basis in momentum space, $i$ is the band index and $q$ is the quasi-momentum, $c_{i,l}(q)$ is the superposition coefficient with $l=0,\pm1,\pm2,...$. While the target state is considered to be the superposition of Bloch states, 
$$\psi= \sum_{i}A_{i}|i,q\rangle=\sum_{i}A_{i}\sum_{l}c_{i,l}(q)|2l\hbar k+q\rangle$$ 
with $A_{i}$ is the superposition coefficient.

Varying the pulse parameters, we numerically find a time sequence 
$$\{t_1=0, t_2, t_3, ...\}$$ 
of optical lattice pulses 
$$ V_{latt}(t)=V_0 \ \mathrm{for} \ (t_{2i-1} < t < t_{2i})$$  
$$V_{latt}(t)=0 \ \mathrm{ for } \ (t_{2i} < t < t_{2i+1})$$ 
that creates  a $\frac{\pi}{2}$ pulse in the Ramsey interferometer, i.e., after this operation, the atoms on S-band and D-band are transferred into target states $(|S\rangle+|D\rangle)/ \sqrt{2}$ and $(-|S\rangle+|D\rangle)/\sqrt{2}$, respectively. In our designed pulse sequences, both the laser intensity and the duration can be changed. In the real experiment, we keep the lattice depth constant and only change the time lattice on and off for experimental simplicity without losing of fidelity.

The evolution operator for ${t_{2i-1} < t < t_{2i}}$ is 
$$\hat{U}_\mathrm{on}(t_{2i}-t_{2i-1})=\exp\{-\frac{i}{\hbar}\int_{0}^{t_{2i}-t_{2i-1}} \lbrack\frac{p^2}{2m}+V_{0}\cos^2(kx)\rbrack dt\}$$  and 
$$\hat{U}_\mathrm{off}(t_{2i+1}-t_{2i})=\exp(-\frac{i}{\hbar}\int_{0}^{t_{2i+1}-t_{2i}}\frac{p^2}{2m}dt)$$ 
for the free evolution operator at ${t_{2i} < t < t_{2i+1}}$, where $p$ and $m$ are respectively the momentum and mass of a single atom. After applying the pulse sequence, the final states $|\psi_{f_1}\rangle$ and $|\psi_{f_2}\rangle$ can be written as
\begin{eqnarray}
|\psi_{f_1}\rangle &=& \prod_{i} \hat{U}_\mathrm{on}(t_{2i}-t_{2i-1}) \cdot \hat{U}_\mathrm{off}(t_{2i+1}-t_{2i}) \cdot |S\rangle \\
|\psi_{f_2}\rangle &=& \prod_{j} \hat{U}_\mathrm{on}(t_{2i}-t_{2i-1}) \cdot \hat{U}_\mathrm{off}(t_{2i+1}-t_{2i}) \cdot |D\rangle.
\end{eqnarray}

By changing the parameters of the pulse sequence, one can maximise the two fidelities $F_{1}$=$|\langle \psi_{A_1} | \psi_{f_1}\rangle|^2$ and $F_{2}$=$|\langle \psi_{A_2} | \psi_{f_2} \rangle|^2$. For $V_0=10E_\mathrm{r}$ the time sequence for $\pi/2$ pulse as $\{0, 56.2, 84.2, 106.8, 129.9\}(\mu$s) with $F_1=96.9\%,F_2=97.9\%$.

The method for designing $\pi$ pulse is similar to the case of $\pi/2$ pulse, but with other target states. After the sequence of optical lattice pules, the atoms should be transferred from $|S\rangle$ or $|D\rangle$ into the target states $|D\rangle$ or $-|S\rangle$, respectively. For $V_0=10E_\mathrm{r}$ the time sequence for $\pi$ pulse as $\{0, 49.2, 101.7, 123.8, 150.2\}(\mu$s), and the transfer of $|S\rangle$ and $|D\rangle$ to the target states are with the fidelity $98.5\%$ and $98.0\%$, respectively.

During design of the pulse sequences, we have neglected atom-atom interaction because the duration of pulses is short enough and atom-atom interaction strength in our system is weak~\cite{Liu}. Numerical results of the GPE with atom-atom interaction show that the interaction leads to a change of fidelity less than $1\%$ compared with our designed time sequence.

\section{Data analysis}

After the band mapping, we release the condensate from the trap and let it expand freely for 31 ms to perform time of flight (TOF) imaging. The TOF image shows the atoms from the S band distributed in the center $0\hbar k$ while atoms from the D-band are distributed on the side zone $\pm2\hbar k$. 
We apply an algorithm to remove the background fringes~\cite{fringe}, and integrate the atom distribution in $y$-direction of the TOF image.
Then the atomic distribution is fitted with a bimodal function
\begin{eqnarray}
	f(x)&=&\sum\limits_{i=1,2,3}G_iF_{th}(e^{-(x-x_i)^2/\sigma^2}) + \nonumber \\ 
	& & + \sum\limits_{i=1,2,3}H_i(\max[1-(x-x_i)^2/\chi_i^2,0])^2
\end{eqnarray}
with a Gaussian thermal distribution $F_{th}$ and a Thomas Fermi like the condensate. $i=1,2,3$ corresponds to the three atom clouds for $-2\hbar k$, $0 \hbar k$, and $2\hbar k$, respectively.
We also add a restriction to all six amplitude terms $H_i$ and $G_i$ as $H_1/G_1=H_3/G_3$. From the amplitude of each component $H_i$ and the width of condensate part $\chi_i$, we can determine the population of condensate in the S-band $N_S$ and in the D-band $N_D$, and further calculate the population of the D-band as $p_D(t_{OL})=N_D/(N_S+N_D)$. The experimental contrast at a certain time is given by fitting the measured amplitude of oscillation $p_D$ for 100$\mu s$ (about 2.5 periods) with a time step 5$\upmu$s using a cosine function, where the errorbar is given by the fitting error with 95$\%$ confidence bounds.

\section{Calculation of contrast decay at zero temperature within an echo-Ramsey interferometer}

\subsection{Contrast decay induced by atom-atom interaction} 

When atoms collide in the excite band they can decay to a lower band ~\cite{Spielman06,Bucker11}. We calculate the collision decay of atom population in different bands using a simple rate equation model ~\cite{Zhai},
%model. The decreasing of the atom number $N_{S(D)}(t)$ of the ultracold bosonic atoms in the D-band can be described by the master equation~\cite{Zhai},
\begin{equation}\label{e5}
\frac{dN_{S(D)}(t)}{dt}= -K_{S(D)}N_{S(D)}(t)^2,
\end{equation}
where the factor $K_{S(D)}$ is given by summation of cross section of the two-atom inelastic collision from different channels.
Solving Eq.(\ref{e5}), the population from the D-band $N_D(t)$  decays as $1/(1+t/\tau_D)$, with the time constant of atom decay from the D-band given by $1/\tau_D=K_DN_D$.  $N_S(t)$ can similarly be given from the collision rate of atoms between the S- and the D-bands.

We measure the collisional decay rate of atoms in different bands by band mapping without the final $\pi/2$ pulse and get $\tau_D = 1.9 \pm 0.3$ ms, $\tau_S = 5.1 \pm 0.8$ ms. 

%
%\noindent\rd{what is the time constant, give number} \\
%\bl{you write: we used the measured collisional decay of atoms in different bands and get $\tau_D = 1.9 \pm 0.3$ ms, $\tau_S = 5.1 \pm 0.8$ ms }  \\
%\rd{i am not happy with this. this is confusing (how do you get $\tau_S$) and may be even wrong when written and does not sound trustworthy when saying we get a coherence time of 14 ms that is 8 decay times $\tau_D$.  \\
%If we take the period of $\pi$-pulses as $\tau_D$, then for 5 $\pi$-pulses = 1.9*6 ms experiemnt time the remaining atom that had not experienced a decay would only be 12.5\%.  that does not make sense}
%

During the collision decay, the potential energy of the D-band is transferred to kinetic energy in radial direction. This energy is large enough so that the majority of the decayed atoms will not be counted in for the condensed peaks. After the population decay atom number $N_{S}(t)$ and $N_{D}(t)$ is unbalanced which leads to a reduction of the contrast by a factor $\frac{2\sqrt{N_SN_D}}{N_S+N_D}$. From the measured $N_S$ and $N_D$ we find that the influence of this population unbalance to the calculated contrast $C_0(t_{OL})$ is less than $1\%$ for an experiment with $t_{OL} < 2$ ms as shown as the brown dashed line in Fig.3(a). %There\marginnote{{\footnotesize \rd{not clear, what do you want to say here}}} the \gr{influence with the real fidelity} is also included, these two effects would lead to a decrease of contrast by less than $3\%$ for our experiment condition.

Applying the repeated $\pi$-pulses the coherence time is much longer than this decay time: (\emph{i}) The $\pi$-pulses reverse the population in the S- and D-bands, and therefore prevent a strong imbalance. (\emph{ii}) In the data analysis extract the remaining condensed part by bimodal fitting. Both together dramatically reduce the  effect of collisional decal on the contrast of the observed interference between the remaining coherent parts, and the collisional decay time does not limit the coherence time of our interferometer signal, as long as the remained population in condensed part is large enough for detection.

\subsection{The contrast decay induced by the quasi-momentum distribution and the nonuniform of optical lattice potential in radial direction} 

For these we have to turn to numerical calculations taking the real potentials and their variation into account. Considering that the trapping frequencies in $y,z$ direction are very similar in our system, we can use use a mean field model (GPE) in cylindrical coordinates to describe the system at zero temperature. The evolution of the wave function $\Phi(\textbf{r},t)$ is governed by
\begin{equation}\label{e6}
i\hbar \frac{\partial}{\partial t}\Phi(\textbf{r},t)=[-\frac{\hbar^2}{2m}\triangle+V_{\mathrm{ext}}+NU_0|\Phi(\textbf{r},t)|^2]\Phi(\textbf{r},t),
\end{equation}
where $\textbf{r}=(x,r,\theta)$, $V_{\mathrm{ext}}$ is the external potential, and $U_0=\frac{4\pi\hbar^2 a_s}{m}$ is the interaction term with $a_s$ the scattering length.

In our case the radial part of wavefunction is in the ground state and uniform in $\theta$ coordinate.
The lattice potential itself depends on the radial position $r$. We first neglect the kinetic term in radial direction and separate the wavefunction into radial part and axial part as $\Phi(\textbf{r},t)= \frac{1}{\sqrt{2\pi}}\psi(x,t)\phi(r)$, then Eq.(\ref{e6}) can be simplified to the following 1d GPE at a certain value $r=r_i$ as
\begin{equation}\label{e7}
i\hbar \frac{\partial}{\partial t}\psi(x,t)=[-\frac{\hbar^2}{2m}\frac{\partial^2}{\partial x^2}+V_{\mathrm{ext,r_i}}(x,t)+\rho_{r_i}U_0|\psi(x,t)|^2]\psi(x,t)
\end{equation}
$V_{\mathrm{ext,r_i}}$ is the combination of both harmonic potential $V_{\mathrm{trap}}=\frac{m}{2}\omega_x^2x^2$ and lattice potential $V_{\mathrm{latt},r_i}=V_{0}Q(t) e^{-2r_i^2/w_{\mathrm{latt}}^2}\cos^2(kx)$ where $Q(t)$ takes value 0 or 1 depending on the time sequence, and $w_{\mathrm{latt}}$ is the waist of lattice laser. 
$\rho_{r_i}=\frac{\exp[{-m\omega_{r}r_i^2/\hbar}]}{\sqrt{\pi\hbar/m\omega_r}}N$
is the linear density. We solve Eq.(\ref{e7}) to get the $\psi(x,t)$ and the population in the D-band $p_{D_i}$ for position $r_i$. We then calculate 30 different radial positions and take their weighted average according to the atom number distribution $\sum\limits_{i} (2\pi r_i \rho_{r_i}\cdot p_{D_i})/\sum\limits_{i} (2\pi r_i \rho_{r_i})$. Finally the oscillation amplitude of the average $p_D$ is fitted to get a contrast as shown in dotted line in Fig.3(a).

In this simulation, we consider the wave-function's distribution instead of using a single atom model, thus the influence of quasi-momentum distribution is also included automatically.

\subsection{Contrast decay induced by radial motion of the condensate}

In the above calculation, we consider the distribution of lattice potential in radial direction, however, the radial size of he wavefunction also changes with time. To consider this effect we need to estimate the expansion speed of the atom cloud in the transverse direction.

We can take each site of the lattice as a small independent BEC with about $1.5\times 10^3$ atoms (about 65 lattice sites), and calculate how it spreads after the trapping potential changed during switch on and loading into the lattice. When the lattice is turned on, the trapping frequency in x direction increases to $20 kHz$ for 10$E_\mathrm{r}$ which is much larger than the
harmonic trap of $2\pi\times 24 Hz$, this sudden increase of trap frequency induces the spread in the radial direction, which can be calculated as~\cite{Castin},
\begin{equation}\label{e8}
 \ddot{\lambda_x}=\frac{\omega_x^2(0)}{\lambda_x^2\lambda_r^2}-\omega_x^2(t)\lambda_x,\ \ \
 \ddot{\lambda_r}=\frac{\omega_r^2(0)}{\lambda_x\lambda_r^3}-\omega_r^2(t)\lambda_r,
\end{equation}
where $\omega_x, \omega_r = \sqrt{\omega_y \omega_z}$ are the frequency of the effective trapping potential in $x$  and $r$ direction for the small BECs in each lattice site, respectively, and $\lambda_{r}=
r_i(t)/r_{i}(0)$ is the expansion with $r_i(0)$ the initially radial position.  Using this time dependent radial expansion we then follow a procedure like above to calculate the average $p_D$ and contrast as shown in dashed line in Fig.3(a) of the main text.

\subsection{Contrast decay induced by laser intensity fluctuations}

The laser intensity in our experiment fluctuates by about $0.1\%$ during the holding time. The laser intensity changes are slow and do not cause excitation between the different bands. Simulations like above are then repeated with different laser intensities sampled from the measured fluctuations. The averaged result is shown in Fig.3 of the main text with the purple dash-dotted line.   

%In the simulation discussed above, we solve Eq.(\ref{e7}) with a constant lattice depth during the holding time. In fact the laser intensity in our experiment have a fluctuation of about $0.1\%$ during the holding time. We measured the laser intensity and find that it changes slow and would not cause excitation between different bands. Laser fluctuation in 5 different real experiments are measured and the measured fluctuation are taken into 5 separate simulations. Then the averaged result is shown in Fig.3 of the main text with the purple dash-dotted line.

\subsection{Calculation of contrast decay induced by thermal fluctuations }

To calculate that contrast decay induced by thermal fluctuations we apply the finite-temperature truncated Wigner method~\cite{Temperature}. We solve the 1d GPE with a stochastic initial wavefunction $\psi^{\prime}(x)=\psi+\sum_j\psi_j$, where $\psi$ is the zero-temperature condensate wave-function within the one-dimensional optical lattice, $\psi_j$ corresponding to the thermal fluctuations that is given by $\psi_j=A(r_0,\theta_0)[u_j(x)\beta_j-v_j^*(x)\beta_j^*]$, where $A(r_0,\theta_0)=\sqrt{n_{0}(0,r,\theta)/\iint n_0(0,r,\theta)r \mathrm{d}r\mathrm{d}\theta}$, with the condensation density $n_0(x,r,\theta)$. For simplicity and without loss of generality, we choose $r_{0}=0, \theta_{0}=0$. Here $\beta_j$ is a complex number with random phase which satisfies $\beta_j^*\beta_j=\bar{N}_j+1/2$, where the quantum fluctuations are included by the $1/2$ term. in which $\bar{N}_j=\sum_i 1/(e^{E_{ji}/k_BT}-1)$ and $E_{ji}=E_j+\Lambda_i$ is the summation of the energy of the j-th Bogoliubov mode $E_j$ plus the i-th eigen-energy $\Lambda_i$ of the radial harmonic potential. $u_j$ and $v_j$ are the $j$-th Bogoliubov modes solved from the one-dimensional Bogoliubov de Gennes equation as ~\cite{Temperature},
\begin{align}
 [H_{sp}+2U_0n_0-\mu]u_j-U_0n_{0}v_j=E_{j}u_j,\\
  -[H_{sp}+2U_0n_0-\mu]v_j+U_0n_{0}u_j=E_{j}v_j,
\end{align}
with $H_{sp}=-\frac{\hbar^2}{2m}\frac{d^2}{dx^2}+\frac{1}{2}m\omega_x^2x^2$.

In the calculation we use 300 excitation modes and repeat our simulation 15 times with different stochastic initial states. From the wavefunctions and the population distributions, we then obtain the contrast $C(t_{OL})$, calculated at the different holding times $t_{OL}$, which can then be compared to the experiment.  In a separated calculation, we also calculate for the quantum fluctuations by taking $\bar{N}_j=0$, the result is shown in Fig. 3(a) of the main text with green dashed line, which is close to the result without quantum fluctuations, where the coherent time is only reduced by $0.2\%$.

% Furthermore, we fitted the ratio of two contrasts with or without the temperature effect $C/C_0$ with $\exp[-\beta t_{OL}]$, the average and deviation of $\beta$ also can be given.

%\end{spacing}%
%\begin{spacing}{2.0}

%\newpage

%\begin{addendum}
%
%\item [Acknowledgement] We thank Z. K. Chen for the calculation
%of pulse time sequences and discussion. Thank P. Zhang, Q. Zhou,
%and B. Wu for helpful discussion. This work is supported
%by the National Key Research and Development Program of China (Grant
%No. 2016YFA0301501), and the NSFC (Grants No.11334001, No.61475007 and
%No. 61727819). G. J. Dong acknowledges the support by the National
%Science Foundation of China (Grants No. 11574085 and No. 91536218),
%and 111 Project ( B12024).  
%JS acknowledges support by the European Research Council, ERC-AdG QuantumRelax. 
%
%\item [Author Contributions] All the authors contributed to the
%work and helped in editing the manuscript.
%
%\item [Competing Interests] The authors declare that they have no
%competing financial interests.
%
%\item [Correspondence] Correspondence should be addressed to
%Xiaoji Zhou(xjzhou@pku.edu.cn) and J. Schmiedmayer
%(schmiedmayer@atomchip.org).
%\end{addendum}

%\end{spacing}

\end{document}